\documentclass[twocolumn,english,aps,prb,floatfix,preprintnumbers,showpacs,amsfonts,amssymb,superscriptaddress]{revtex4}
\usepackage{graphicx}
\usepackage{dcolumn}
\newcommand{\mypath}[1]{./#1}

\begin{document}

\title{Colossal Magnetoresistance Observed in Monte Carlo Simulations \\
of the One- and Two-Orbital Models for Manganites}

\author{Cengiz \c{S}en}
\affiliation{National High Magnetic Field Laboratory and Department of Physics,
Florida State University, Tallahassee, FL 32310}

\author{Gonzalo Alvarez}
\affiliation{Computer Science \& Mathematics Division, 
Oak Ridge National Laboratory, Oak Ridge, TN 37831}

\author{Horacio Aliaga}
\affiliation{Condensed Matter Sciences Division, Oak Ridge National Laboratory, Oak Ridge, TN 32831}
\affiliation{Department of Physics and Astronomy, The University of Tennessee, Knoxville, TN 37996}

\author{Elbio Dagotto}
\affiliation{Condensed Matter Sciences Division, Oak Ridge National Laboratory, Oak Ridge, TN 32831}
\affiliation{Department of Physics and Astronomy, The University of Tennessee, Knoxville, TN 37996}

\date{\today}

\begin{abstract}
The one- and two-orbital double-exchange models for manganites 
are studied using Monte Carlo computational techniques 
in the presence of a robust electron-phonon coupling
(but neglecting the antiferromagnetic exchange $J_{\rm AF}$ between the localized spins).
The focus in this effort is on the analysis of charge transport. Our results for
the one-orbital case confirm and extend previous recent investigations
that showed the presence of robust peaks in the resistivity vs. temperature curves for this 
model. Quenched disorder substantially enhances the magnitude of the effect, while magnetic
fields drastically reduce the resistivity. A simple picture
for the origin of these results is presented.
It is also shown that even for the case of just one electron,
the resistance curves present metallic and insulating regions by varying the temperature, as it
occurs at finite electronic density. Moreover, in the present study these 
investigations are extended to the more realistic two-orbital model for manganites. 
The transport results for this model show large peaks in the
resistivity vs. temperature curves, located at approximately the Curie
temperature, and with 
associated large magnetoresistance factors.
Overall, the magnitude and shape of the effects discussed here closely resemble experiments for
materials such as $\rm La_{0.70} Ca_{0.30} Mn O_{3}$, and they are
in qualitative agreement with the current predominant theoretical view that competition 
between a metal and an insulator, enhanced by quenched disorder, is crucial 
to understand the colossal magnetoresistance (CMR) phenomenon.
In spite of this success, it is argued that
further work is still needed to fully grasp the experimentally observed CMR effect, since in
several other Mn oxides an antiferromagnetic charge-ordered orbital-ordered state is the actual competitor 
of the ferromagnetic metal.

\end{abstract}

\maketitle

\section{Introduction}

One of the most outstanding open problems in the area of transition metal oxides
is the explanation of the colossal magnetoresistance (CMR) effect that appears in the Mn oxides
that are widely referred to as manganites. These compounds present a rich phase diagram
with a variety of competing states which are stabilized by changing the carrier
concentration using a standard chemical doping process involving ions with different valences, 
or by varying the carrier bandwidth via isovalent doping \cite{re:tokura00,re:rao98,re:dagotto05,re:moreo99b,re:ibarra99,re:deteresa97,re:lynn96,re:moreo99,re:uehara99,re:dagotto01,re:salamon01,re:parisi01,re:louca97,re:mathur03,re:yunoki98b,re:yunoki98,re:millis95,re:millis96,re:khomskii01,re:dagotto02,re:verges02,re:ahn04,re:dagotto05b,re:motome03c,re:motome00,re:motome99,re:mathieu04,re:salafranca06}. 
Notorious among the low-temperature regimes
stabilized in manganites
are a ferromagnetic (FM) metallic phase and several antiferromagnetic/charge/orbital ordered insulating
states. For the compounds with intermediate
or small Curie temperatures, the experimentally obtained resistivity vs. temperature
curves present a sharp peak, which occurs precisely
at the transition toward ferromagnetism. In the vicinity of this peak, the CMR effect is observed,
which consists of enormous changes in the resistivity upon the introduction of
relatively small magnetic fields. Although technological applications of CMR compounds in the
read-sensor industry will still need an increase by at least a factor two of the currently available
critical temperatures where the large magnetoeffects occur, the physics behind this
remarkable CMR phenomenon defines a challenging basic-science problem that has attracted
the attention of the condensed matter community.

The explanation of the CMR effect is certainly the crucial goal of theoretical investigations
in the manganite context. Early theoretical work showed that the standard double
exchange (DE) model was not sufficient to understand these materials \cite{re:millis95}.
In fact, a DE model cannot even produce an insulator at high temperatures, in the realistic regimes
of electronic densities \cite{re:calderon99}, and this pointed toward the
importance of other couplings, 
such as electron-phonon, for a proper description of these compounds \cite{re:millis96}.
Progress was later made with the realization that manganite models
have tendencies toward mixed phase regimes, typically involving metallic and
insulating states in coexistence \cite{re:moreo99b,re:dagotto02}. This discovery was possible only after the DE model
and its close variations were studied with unbiased computational methods beyond mean-field approximations.
Inhomogeneous states with a variety of length scales 
appear frequently in these studies and the full strength of computational techniques is clearly
needed to fully understand this and other families of complex oxides \cite{re:dagotto05}. 
The theoretical discovery of phase separation tendencies \cite{re:yunoki98b}  
triggered an enormous experimental effort
that confirmed the relevance of mixed states in most of the CMR compounds 
(for a review see Ref.~\onlinecite{re:dagotto01}). 
Percolative pictures were envisioned to understand these materials. Model calculations by
Mayr {\it et al.} \cite{re:mayr02} and Burgy {\it et al.} \cite{re:burgy01,re:burgy04}, using simplified
spin systems and random resistor networks, revealed a phenomenology very similar
to that of real CMR materials in the regime of couplings and electronic densities where
metallic and insulating states were in competition. In fact, a robust peak in the resistivity of the
resistor network was found at intermediate temperatures, and a huge change in its
value was found to occur in the presence of magnetic fields. The key role of quenched
disorder was remarked in these investigations, to obtain large enough effects \cite{re:burgy01,re:burgy04}.

This initial effort using simple models was followed by
calculations of resistivities in the more realistic, although still simplified, one-orbital model for manganites.
Verges {\it et al.} \cite{re:verges02} numerically showed that an insulator can appear at intermediate and large
temperatures if the electron-phonon coupling $\lambda$ is robust enough. This regime,  caused by
localized polarons, is followed by a rapid transition to a metal upon cooling. 
In this model, the two tendencies in competition at low temperatures are $both$
ferromagnetic, and they only differ in the character of the
charge distribution (uniform vs. localized). Although having two competing FM
states cannot solve the entire CMR issue,  since often the competition in experiments
is between a ferromagnet against an antiferromagnetic/charge/orbital ordered state,
the results were sufficiently interesting and challenging that they deserved further work. Moreover, they could
be of relevance to important Mn oxides such as $\rm La_{0.7} Ca_{0.3} Mn O_3$, which at least
naively seem well separated from charge ordered states in the phase diagrams.
Recently, Kumar and Majumdar \cite{re:kumar06} made a very important contribution by 
proposing a new Monte Carlo (MC) algorithm to study
fairly large lattices of the one-orbital model including phonons. Their main observation is that the clean
limit results of Ref.~\onlinecite{re:verges02} are much enhanced by including  on-site quenched
disorder, together with a robust electron-phonon coupling \cite{re:kumar05}. 
This role of disorder to trigger polaron formation in systems with strong
electron-lattice coupling
is an effect that complements the nanoscale phase coexistence near
a first-order transition boundary also triggered by quenched disorder
emphasized in other studies \cite{re:burgy01,re:burgy04}.
Large peaks in the
resistivity vs. temperature curves were reported in Ref.~\onlinecite{re:kumar06}, resembling experiments for some 
manganites. Their conclusion regarding the importance of
disorder was in agreement with previous
investigations  \cite{re:burgy01,re:akahoshi03,re:nakajima02,re:burgy04,re:kumar04},
and provided further confirmation of the currently widely accepted view of manganites, namely that the
essence of CMR lies in the competition of phases (metal vs. insulator), 
supplemented by quenched disorder 
to obtain large enough effects in sufficiently wide regions of parameter space.

In spite of this tremendous progress, there are still several aspects of the CMR problem that need
further refinement. Two issues are notorious: (i) It is important to show that the
results previously obtained for the one-orbital model, focusing on the resistivity peak, 
do also appear for a more realistic
two-orbital model. Several manganites present orbital order and, as a consequence, using two orbitals
per Mn atom is crucial for a proper description of these materials; (ii) The consideration 
of the antiferromagnetic spin coupling $J_{\rm AF}$ between the localized $t_{\rm 2g}$ spins is also crucial.
For instance, this coupling is needed to stabilize several important 
phases with charge/orbital and antiferromagnetic order, as previously
shown \cite{re:dagotto01,re:hotta00}. 
The full understanding of the CMR effect needs these two extra refinements.

In the present paper, investigations of both the one- and two-orbital models for manganites
are reported, with emphasis on the resistivity vs. temperature curves. The main results discussed
in this paper are the following: {\it (1)} We provide an independent confirmation of the results 
of previous investigations discussed in  Refs.~\onlinecite{re:verges02} and \onlinecite{re:kumar06}.
The study of resistivity in Ref.~\onlinecite{re:kumar06} relied on the analysis of the 
optical conductivity extrapolated to zero frequency.
In addition, a novel algorithm was used to generate the classical spin configurations. In our present effort,
a different numerical method (exact diagonalization) 
is used and, more importantly, the transport properties are
estimated using the Landauer formalism based on transmission coefficients. Fortunately,
our study shows that the results of Kumar and Majumdar \cite{re:kumar06} reporting robust resistivity
peaks in the one-orbital model are indeed qualitatively correct, even when a fairly different approach is used to
calculate transport properties. This confirmation of previous investigations helps providing a robust
foundation to computational studies of models for manganites. 
{\it (2)} Still within the one-orbital model context,
here a comprehensive analysis of the influence of quenched disorder and electronic density is provided, thus
considerably extending the studies reported in previous efforts.
A surprising result is that even just one $e_{\rm g}$ electron on a finite lattice
can present transmission characteristics that include a resistance vs. temperature curve in qualitative agreement
with results at finite electronic densities $n$ and with experiments. Charge localization is found to be
responsible for all these features, as previously remarked in Refs.~\onlinecite{re:verges02},\onlinecite{re:kumar06}
as well. A toy example is here discussed to understand these results in very simple terms.
{\it (3)} Finally, one of our main new contributions is the extension of the previous investigations in the
one-orbital model context into a two-orbital model framework. After a comprehensive analysis of the two-orbital
model properties, reported here, it is concluded that this 
model presents a phenomenology similar to that of the one-orbital model simulations, at least for the
case $J_{\rm AF}$=0. In other words, sharp peaks in the resistivity vs. temperature 
are observed in robust regions of parameter space. This conclusion adds more evidence
that theoretical investigations are on the right track toward an understanding of the challenging
CMR effect. 

However, the important inclusion of $J_{\rm AF}$ is postponed for future investigations. Working at
the special case $J_{\rm AF}$=0 much simplifies the numerical analysis, particularly regarding the
convergence properties: the localized spins do not have conflicting tendencies, such as ferro- and
antiferro arrangements with close energies, thus they rapidly tend toward ferromagnetic ground states 
at low temperatures, even if the initial starting Monte Carlo configuration is random.
The assumption $J_{\rm AF}$=0 effectively reduces the global effort to merely making sure 
that the classical lattice displacements regulated by the electron-phonon coupling are properly
converged. Important metastabilities were not found in our investigations.
The considerably more subtle technical difficulties that will arise with the
inclusion of $J_{\rm AF}$ in the model Hamiltonians are left for future considerations.

The organization of the paper is the following. In Section II, results for the one-orbital model are presented,
starting with a brief discussion of the model itself and technical aspects.
The main portion of this section is devoted to the numerically calculated
resistivity vs. temperature curves, obtained at several electronic densities,
electron-phonon couplings, and strength of the quenched disorder. The results for the one electron problem
are included in this section, together with evidence that charge localization is responsible for the insulating
regime. A simple toy example is presented to understand the results. 
In Section III, a similar analysis is presented but using the two-orbital model. Section IV
contains the conclusions of our effort, and suggestions for further work.

\section{ONE-ORBITAL MODEL}\label{sec:oneband}

\subsection{Definition and Details of Simulation}

The one-orbital model used in this study is given by:
\begin{eqnarray}
H_{1b}&=&-t\sum_{\langle ij \rangle , \alpha} (c_{i,\alpha}^{\dag}c_{j,\alpha} + \mbox{h.c.})
-J_{\rm H}\sum_{i,\alpha, \beta}c_{i,\alpha}^{\dag} {\vec \sigma}_{\alpha,\beta}c_{i,\beta}\cdot {\vec S}_i \nonumber
\\ &-&\lambda t \sum_{i,\gamma,\alpha}(u_{i,-\gamma}-u_{i,\gamma})c_{i,\alpha}^{\dag}c_{i,\alpha}
+t\sum_{i,\gamma}(u_{i,\gamma})^2 \nonumber \\ &+& \sum_{i,\alpha}
(\Delta_{i}-\mu) n_{i,\alpha},
\end{eqnarray}
where $c_{i,\alpha}^{\dag}$ creates an electron at site $i$ with spin $\alpha$,
$\mathbf{\sigma}_{\alpha,\beta}$ are the Pauli spin matrices, $\langle ij \rangle$ indicates 
summing over nearest neighbor sites, and $t$ is the nearest neighbor 
hopping amplitude for the movement of electrons ($t$ also sets the energy unit, i.e. $t$=1
in all of the results below).
The first and second terms are the standard for a double exchange model, with ${\vec S}_i$
being a classical localized spin that represents the $t_{\rm 2g}$ degrees-of-freedom. 
The third term in the Hamiltonian accounts for the energy corresponding 
to the lattice-carrier interaction, with $\lambda$ being the strength of the electron-phonon 
coupling. $u_{i,\gamma}$ 
are the distortions (lattice displacements) of the oxygen atoms 
surrounding a Mn ion at site $i$. The index $\gamma$ in 3D (2D) runs over 
three (two) directions $x,y$ and $z$ ($x$ and $y$). The tendency toward increasing the
magnitude of the lattice distortions is balanced by the fourth term in the Hamiltonian, which represents 
the stiffness of the Mn-O bonds. Since the study of quantum phonons in this context is not
possible with currently available algorithms, 
the oxygen displacements are considered classical, approximation widely used in studies
of manganites \cite{re:dagotto01}. Finally, the last term corresponds 
to the quenched disorder, which here it is introduced in the form
of random site energies. $\Delta_i$ represents the strength of the
disorder at a given site, and these numbers
are chosen from a bimodal distribution of width $2\Delta$ with mean 0.
The overall electronic density $n$ 
is controlled with the help of a chemical potential $\mu$ 
added to the last term in the Hamiltonian. 
In the rest of the paper, for simplicity spatial labels 
will be denoted without arrows or
bold letters independently of the dimension. Also the notation $i+j$ is meant to
represent the lattice site given by the vectorial sum of the vectors
corresponding to $i$ and $j$, respectively. 

In this manuscript, the limit of an infinite Hund coupling will be considered, which is another widely used
simplification known to preserve the essential physics of manganites \cite{re:dagotto01}. 
In this limit, the spin of the 
$e_{\rm g}$-electron perfectly aligns along the localized $t_{\rm 2g}$-spin direction, and the 
Hamiltonian is reduced to:

\begin{widetext}
\begin{eqnarray}
H_{1b} &=& -t\sum_{\langle ij \rangle}\{[\cos\frac{\theta_i}{2}\cos\frac{\theta_j}{2}
+\sin\frac{\theta_i}{2}\sin\frac{\theta_j}{2}e^{i(\phi_{i}-\phi_{j})}]d_{i}^{\dag}d_{j}+\mbox{h.c.}\}
-\lambda t\sum_{i,\gamma}(u_{i,-\gamma}-u_{i,\gamma})d_{i}^{\dag}d_{i}\nonumber \\
&+&t\sum_{i,\gamma}(u_{i,\gamma})^2+\sum_{i}(\Delta_{i}-\mu) n_{i},
\end{eqnarray}
\end{widetext}
where $\theta_{i}$ and $\phi_{i}$ are the spherical coordinates of the core spin at 
site $i$ (assumed classical). The operators $d_{i}^{\dag}$ now create an electron at site $i$ with spin parallel to
the core spin at $i$, and $n_{i}=d_{i}^{\dag}d_{i}$. Note that for an
infinite Hund coupling, the system can be shown to be
particle-hole symmetric with respect to density $n$=0.5. Thus, results
at densities $n$ and 1-$n$ are equivalent.

The technique used here to handle this Hamiltonian involves the standard exact diagonalization of the quadratic 
fermionic sector for a given spin background \cite{re:dagotto01,re:dagotto02}. 
The procedure then consists of an evolution in
 Monte Carlo steps,
where new spin configurations are accepted or rejected according to a standard Metropolis algorithm. 
Details have been widely discussed in previous studies and they will not be 
repeated here \cite{re:dagotto01,re:dagotto02}.
Thermal averages of operators such as spin-spin correlation $\vec{S}_{i}\cdot\vec{S}_{j}$ are 
calculated by carrying out an average over all Monte Carlo steps during the MC evolution, after
discarding the initial set needed to thermalize. Correlation functions at a particular distance
are obtained by averaging over all the possible pairs of sites separated by that distance. As example,
the definition of the spin correlations at distance $x$ is the following:

\begin{equation}
S(x)=\frac{1}{N}\sum_{i}\langle \vec{S}_{i}\cdot\vec{S}_{i+x}\rangle = \frac{1}{N}\sum_{i}
\frac{\mbox{Tr}[\vec{S}_{i}\cdot\vec{S}_{i+x}e^{-\beta H}]}{\mbox{Tr}[e^{-\beta H}]},
\end{equation}
where $\beta$ is the inverse temperature and $N$ is the total number of sites, and the rest of 
the notation is standard. In the one-orbital study, mainly lattice sizes $8\times8$ and $4\times4\times4$ were used. 
In addition, $10^4$ steps were typically employed for thermalization, 
followed by another $10^4$ for measurements. For larger
lattices, such as $12\times12$ and $6\times6\times6$, $10^4$ measurement steps were performed after 
$2,000$ steps for thermalization. Most of the simulations have started with a random configuration
of spins, but simulations with a FM starting configuration have also been carried out in order to check
for convergence. No problems were found in this context, namely both approaches led to very similar
results. Furthermore, independent Monte Carlo runs corresponding to different starting random seeds 
for the initial random spin configuration have also been averaged
wherever possible to increase the accuracy of the results.

The resistivity $\rho$ has been calculated by taking the inverse of the mean conductivity $\sigma$, 
where the latter is related to the conductance $G$ by $G=\sigma L^{d-2}$, with $d$ being the dimension and
$L$ the linear size of the lattice. The calculation of the conductance $G$ has been carried out
following the approach extensively discussed before by Verges \cite{re:verges99}.  The use
of the resistivity notation is to facilitate the interpretation of results and comparison with experiments, namely
we do not
claim to have observed Ohmic behavior in our small system simulations. For the purposes of our paper, whether the
resistivity or resistance is used as the key observable the conclusions are the same. The units used for the
resistivity in the entire manuscript 
are $[h/e^2]$ in 2D, and $[h/e^2]\times~L$ 
in 3D. Precisely in 3D, the results presented in figures
were obtained by multiplying the resistance by the linear size $L$,
assuming a lattice spacing one. To restore the proper units to our
results, the real lattice spacing of Mn oxides must be used.

Finally, it is important to remark that 
a sizable portion of the computational work presented here was carried out
on parallel supercomputers, in particular on the
 NCCS XT3 supercomputer  (2.4-GHz AMD Opteron processor and 2 GB of
memory) at Oak Ridge National Laboratory. 
Typical simulations in this effort made use 
of 100 to 200 nodes in parallel. These
supercomputer resources have decreased substantially the amount of real time
that would have been needed. Indeed, we estimate that the entire effort would
have taken at least one year to complete on standard small-size computer clusters.
The message-passing interface was used to parallelize
the runs that sweep over the various Hamiltonian parameters such as $\lambda$
and temperature. Furthermore, quenched disorder
adds an extra level of computational 
effort since it requires the simulation
and average of results from many different configurations.
This extra level of complexity has also been parallelized.

\subsection{Density n=0.3}

\begin{figure}
\centerline{
\includegraphics[clip,width=8cm]{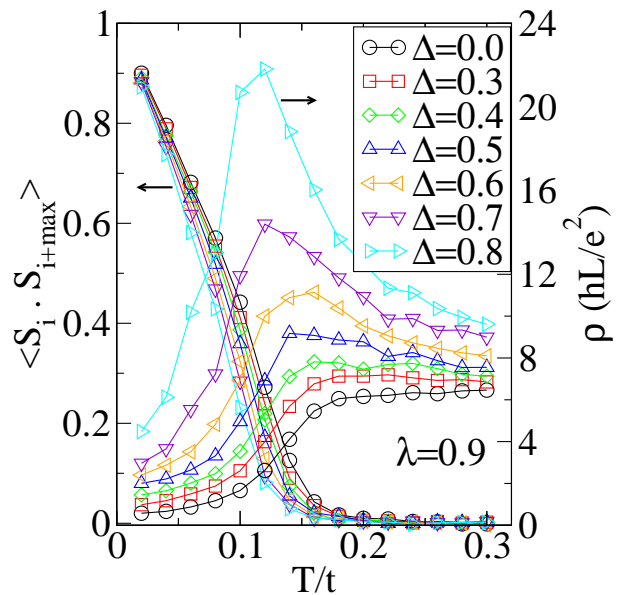}}
\caption{(Color online)  
Monte Carlo results obtained using a $4\times4\times4$ lattice. Shown
are the resistivity and spin-spin correlations, the latter at the maximum allowed distance ($2\sqrt{3}$),
vs. temperature, working at $\lambda$=$0.9$, $n$=0.3, and
for the disorder strengths $\Delta$ indicated. 
The results shown are mainly for one configuration of quenched disorder,
but as many as 10 configurations were used in particular cases of
temperatures and $\Delta$'s, and no
substantial deviations were observed between disorder configurations.
}
\label{Figure1}
\end{figure}

The discussion of our computational results starts at the electronic density $n$=0.3 (equivalent to $n$=0.7,
due to the symmetry discussed in the previous section). Figure \ref{Figure1} is a typical example of the 
resistivity curves
obtained in the present effort. Shown are both the spin-spin correlation at the maximum distance possible in the cluster under study
and the resistivity, working at a fixed electron-phonon coupling $\lambda$=0.9, and varying the strength
of the quenched disorder $\Delta$. In the clean limit, $\Delta$=0, there is a rapid change in resistivity
near the transition to ferromagnetism. This is a typical pure double-exchange behavior: in the
absence of a sufficiently strong $\lambda$, quenched disorder, or other couplings that may lead to 
competing states, then
a metal is obtained at temperatures above the Curie temperature. As already clearly established in this field,
pure double-exchange models are not enough to address the physics of the CMR materials. However, note the
dramatic effect of quenched disorder on the resistivity, 
as shown in Fig.~\ref{Figure1}. As already recently remarked by 
Kumar and Majumdar \cite{re:kumar06}, disorder can induce a peak in the resistivity  that much resembles 
experimental results for typical CMR materials. Even for the small systems studied here, the ratio of resistivities
between its maximum and minimum values 
is as large as $\sim$6 for $\Delta$=0.7. Note the correlation between the peak location and
the temperature where ferromagnetic order appears (signaled in our calculations by the value of
the spin-spin correlation at the largest possible distance in the cluster under investigation). 
The comparison of our 
results with those of recent publications also show that
the different ways to study the resistivity 
in Ref.~\onlinecite{re:kumar06} and here, lead to similar 
qualitative data. This is an important observation, 
in view of the approximate nature of the calculations, that
confirms the results of Ref.~\onlinecite{re:kumar06} and shows that the peak in the resistivity is a
robust feature of the model.

\begin{figure}
\centerline{
\includegraphics[clip,width=6.5cm]{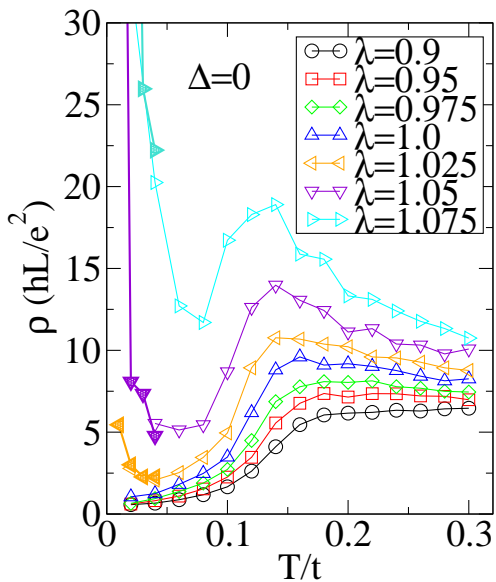}}
\caption{(Color online) Influence of the electron-phonon coupling
$\lambda$ on the resistivity curves in the 
clean limit $\Delta$=0, considering a $4\times4\times4$ lattice, and $n$=0.3.}
\label{Figure2}
\end{figure}

Although a variety of previous theoretical and experimental
investigations have convincingly shown the importance of quenched disorder in the CMR context,
nevertheless it is interesting to observe that a resistivity peak can also be found by varying $\lambda$
even in the clean limit $\Delta$=0, as shown in Fig.~\ref{Figure2}. This observation
will appear repeatedly in the rest of the results discussed below, namely there seems to exist a qualitative
relation between increasing $\Delta$ at small $\lambda$ and simply increasing $\lambda$ at $\Delta$=0.
This fact will be exploited in the studies presented below to simplify our task, 
since a simulation with nonzero quenched disorder needs averages over several disorder configurations,
rendering the effort more time consuming than a clean-limit analysis. However, there is an important
difference between the two cases:
%
observing the resistivity peak in the clean limit requires a $fine$ $tuning$ of $\lambda$.
For a nonzero $\Delta$, the range of couplings with a resistivity peak is much wider (see below
in the $n$=0.1 subsection for a more detailed discussion).
Fine tuning
is not compatible with the CMR effect since the phenomenon appears in a large number of manganese oxides, with
a distribution of $\lambda$'s.
Nevertheless, while it is clear that working at nonzero $\Delta$ and smaller than critical
$\lambda$ is more realistic, to the extent
that  the emphasis of a clean-limit investigation  in a fine-tuned range of $\lambda$
is restricted to the vicinity of the Curie temperature, then both approaches appear to lead to
similar conclusions. Finally, note that at very low 
temperatures the clean limit result shows insulating 
behavior, while the results with a nonzero disorder 
strength do not present such a feature. Although this fact
establishes an interesting difference between the two 
approaches, and in addition it is known that some manganites
do present such an upturn in resistivity 
at low temperatures \cite{re:dagotto02}
the issue will not
be studied in detail in the rest of the manuscript, 
since the focus of the effort is in the resistivity 
peak near the Curie temperature. The analysis of the 
origin of the low-temperature
resistivity upturn in the clean limit is left for future work.

\begin{figure}
\centerline{
\includegraphics[clip,width=6.5cm]{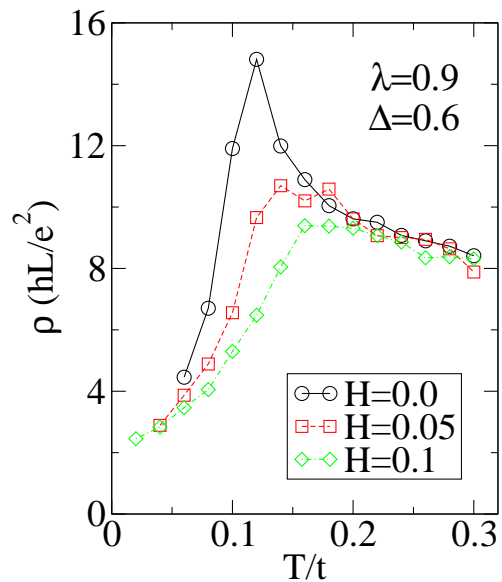}}
\caption{(Color online) Effect of magnetic fields (values indicated)
on the resistivity curves,
for $\lambda$=$0.9$ and $\Delta$=$0.6$, using a $4\times4\times4$ lattice.
The density is $n$=0.3.}
\label{Figure3}
\end{figure}

The resistivity curves with peaks at intermediate temperatures (Figs.~\ref{Figure1},\ref{Figure2})
resemble the experimental data corresponding to 
real manganites. It is remarkable that the numerical results also resemble
the manner in which these curves are affected by magnetic fields. A typical
example is shown in Fig.~\ref{Figure3}, where external fields of a small value 
are used (note that these fields are small when compared
with the natural unit, i.e. the hopping amplitude. However,
in physical units such as Teslas a field $H$=0.1 can be substantial). 
The region affected the most by magnetic fields is the vicinity of the peak. The qualitative
similarity with experiments is obvious, although it is fair to remark that in this subsection a linear scale
is used for $\rho$ while in most of the materials with truly CMR effects a logarithmic scale is needed,
showing that the magnitude of the effect discussed here is substantially 
smaller. This quantitative difference
could be related to the small size of the Monte Carlo
systems used or, more likely, with the absence of a strongly insulating
charge-ordered orbital-ordered antiferromagnetic 
state as the direct competitor of the FM metallic state. 
Here the competitor of the FM metallic state is insulating but also ferromagnetic
and, as a consequence, their resistivities near the Curie transition are not dramatically different. 
Nevertheless, even if only qualitatively, the similarity of the Monte Carlo 
data in Figs.~\ref{Figure1},\ref{Figure2},\ref{Figure3} with experimental observations
is excellent.

\subsection{Density n=0.1}

\begin{figure}
\centerline{
\includegraphics[clip,width=7cm]{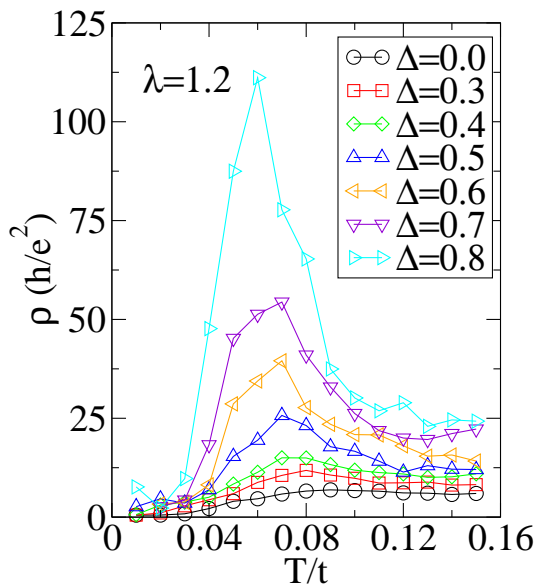}}
\caption{(Color online) Influence of the disorder strength $\Delta$ 
(indicated) on the resistivity vs. temperature curves, working
at $\lambda$=$1.2$, $n$=0.1, and using an $8\times8$ lattice.}
\label{Figure4}
\end{figure}

Several of the effects discussed at the realistic density $n$=0.3 in the
previous section were found to be magnified by reducing the electronic density. 
In this subsection, the case of $n$=0.1 will be considered in detail (as already remarked, $n$=0.9
is the same, via symmetry considerations for the model used). The lattice to be shown is now
two dimensional, to illustrate the similarity systematically found between results in two and three dimensions.
In Fig.~\ref{Figure4}, the influence of the quenched disorder strength $\Delta$ on the
resistivity plots is shown. As in Fig.~\ref{Figure1}, the case of a ``small'' $\lambda$ is
considered first, namely one where in the clean limit the resistivity does not present insulating
behavior. As found for $n$=0.3, with increasing $\Delta$ a prominent peak is generated, which is
located at the Curie temperature (conclusion based on the study of spin correlations, not shown).
The ratio of the maximum and minimum resistivities is now 30-50 in the range of $\Delta$ analyzed here, 
considerably larger than at $n$=0.3.

\begin{figure}
\centerline{
\includegraphics[clip,width=7cm]{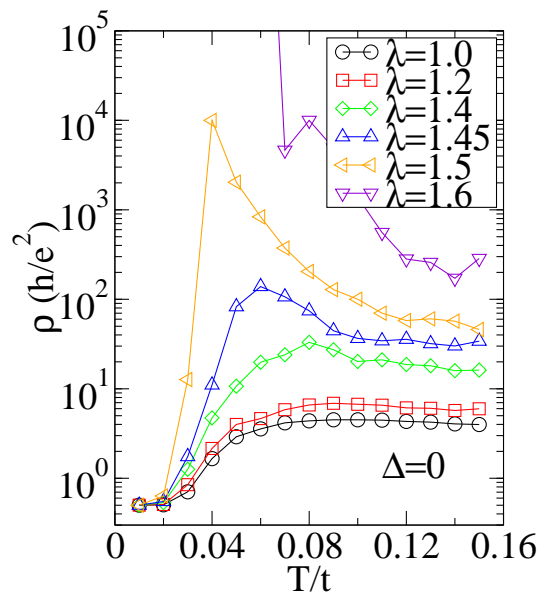}}
\caption{(Color online) Influence of the electron-phonon coupling $\lambda$
on the resistivity vs temperature curves,  in the clean limit  $\Delta$=0, at $n$=0.1, and
using an $8\times8$ lattice.}
\label{Figure5}
\end{figure}

As remarked for $n$=0.3, there appear to exist analogies between the processes of
increasing $\Delta$ at ``small''
$\lambda$ and increasing $\lambda$ in the absence of quenched disorder. This relation
is clear as well at $n$=0.1, and part of the evidence 
is shown in Fig.~\ref{Figure5}, which was obtained in the clean limit. 
In a narrow $\lambda$ range, a prominent
resistivity peak is found, as in Fig.~\ref{Figure2}. Note the use of
a logarithmic scale for the resistivity, showing that the magnitude of the
effect is truly colossal.

\begin{figure*}
\centerline{
\includegraphics[clip,width=15cm]{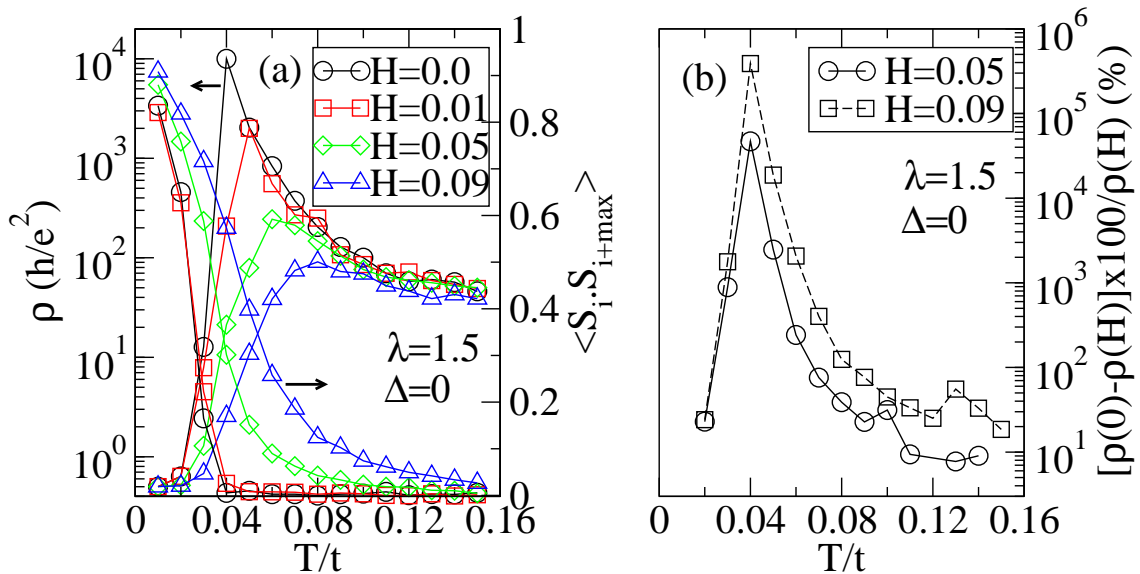}}
\caption{(Color online) (a) Influence of magnetic fields on the resistivity 
curve and on the spin-spin correlation at the maximum allowed
distance ($4\sqrt{2}$) on an $8\times8$ lattice, in the clean 
limit $\Delta$=0. (b) Magnetoresistance ratios vs. temperature, calculated 
for two representative magnetic fields. In both (a) and (b), $\lambda$=1.5 and $n$=0.1.}
\label{Figure6}
\end{figure*}

The effect of magnetic fields at $n$=0.1 is very pronounced (see Fig.~\ref{Figure6}),
once again resembling the magnitude of the CMR effect in real materials. The region in
the vicinity of the resistivity peak is the most affected. The magnetoresistance ratios
(right panel) are as large as those reported in the real Mn oxides with the largest CMR effects. 
The trade-off is that the effect occurs only in a small window of $\lambda$, but this
range, as well as the magnetoresistance value, can be further enlarged by adding quenched
disorder. 

\begin{figure}
\centerline{
\includegraphics[clip,width=8.5cm]{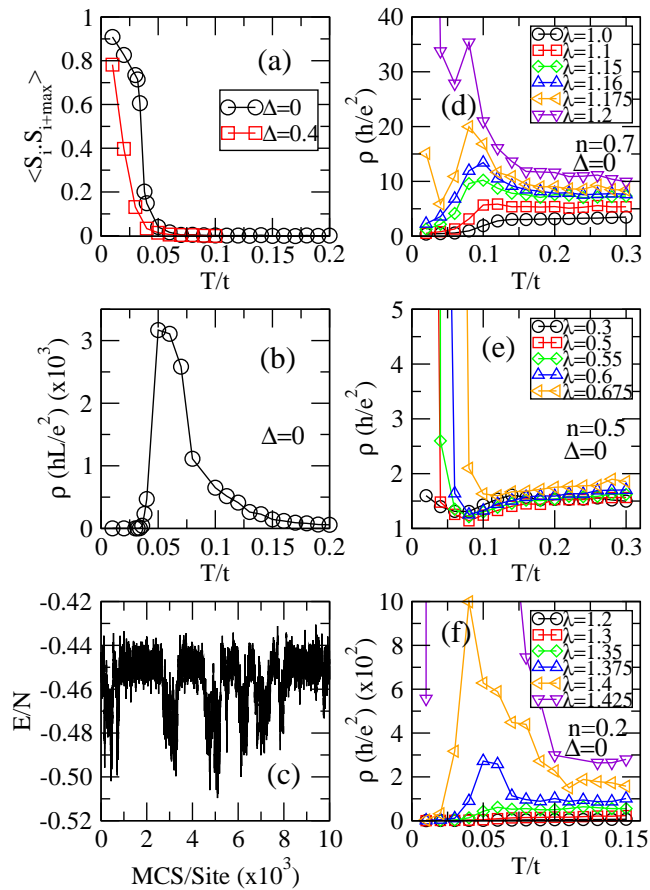}}
\caption{(Color online). Results mainly in the clean limit $\Delta$=$0$
illustrating a variety of issues discussed in the text.
(a),(b), and (c) show the first-order (discontinuous) 
character of the transition
at $n$=0.9, $\lambda$=1.4, using an $8\times8$ lattice. (a) are the spin-spin
correlations at the maximum distance. Also shown in red are results at
$\Delta$=0.4 and averaged over 5 disorder configurations, showing the smearing
of the transition with disorder; (b) is the resistivity vs. $T$;
and (c) is the Monte Carlo time evolution of the energy at $T$=0.034, showing
the presence of two states. (d) is the resistivity vs. temperature
at $n$=0.7 (8$\times8$ lattice), at the $\lambda$'s indicated. 
(e) is the same as (d) but
at $n$=0.5 and using a 12$\times$12 cluster. (f) is the same as (e) but for $n$=0.2.}
\label{Figure7}
\end{figure}

For the particular case $n$=0.1, 
it is interesting to remark the abruptness of
the changes in the resistivity near the peak in Fig.~\ref{Figure5}, 
that resemble a first-order transition
The same occurs at the equivalent density $n$=0.9, as 
shown in Fig.~\ref{Figure7} (a,b,c).
There, the results of a longer Monte Carlo time simulation are presented and these numbers
strongly suggest that indeed a first-order
transition occurs. The evidence is the jump found in (a) the spin correlations and (b) the
resistivity. Also, in (c) the MC time evolution for the energy is shown. This presents sudden
events, that resemble tunneling between two clearly distinct states. Note that  
the temperature chosen
is slightly biased toward the highest energy state, since it is very difficult to fine tune
$T$ such that both competing states are visited an approximately equal amount of MC time.
The first-order nature of the transition also highlights clear similarities with experiments
for some manganites, such as $\rm La_{0.7} Ca_{0.3} Mn O_3$. 
Other Mn oxides appear to have a broader transition.
Nevertheless, it is remarkable that the much simplified one-orbital model used here and in
studies by other groups can be so rich to reproduce even
this type of experimentally observed features.

\begin{figure}
\centerline{
\includegraphics[clip,width=6.5cm]{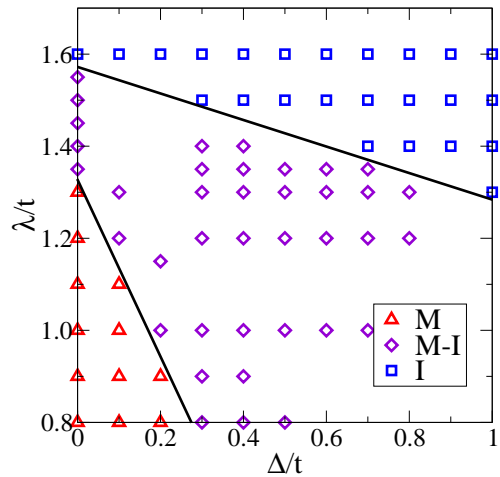}}
\caption{(Color online) 
Influence of quenched disorder on the size of the parameter-space
region where the resistivity peak exists. In the plane $\lambda$-$\Delta$, M (I) denotes
the region where the resistivity is metallic (insulating) at all temperatures, while M-I
is the area where the resistivity peak is present. The calculation was done on an $8\times8$ cluster,
with $n$=0.1.
}
\label{Figurelast2}
\end{figure}

To finalize the study at density $n$=0.1, it is important to address to what extent quenched disorder
(i.e. $\Delta$) does
play a key role in generating the resistivity peak. After all, both in this subsection and at $n$=0.3
it was observed that even  in the clean limit $\Delta$=0 there is a 
$\lambda$ range where a peak is present.
The key observation is that in the clean limit a $fine$ $tuning$ of $\lambda$ is needed to obtain
the resistivity peak, namely the peak only exists in a narrow window of parameters. Including quenched disorder the range becomes much wider. For instance, in
Fig.~\ref{Figurelast2} the area where a resistivity peak exists is shown in the $\lambda$-$\Delta$ plane.
This region rapidly grows with increasing $\Delta$. Avoiding fine tuning of couplings is crucial to
understand CMR materials, since a wide variety of Mn oxides -- with a distribution of bandwidths and
couplings -- present the CMR effects. Any proposed mechanism must be fairly universal to be robust, and the
inclusion of quenched disorder indeed renders the range of couplings for CMR much 
wider than in the clean limit. 

\subsection{Other Electronic Densities \\
and Finite-Size Effects}

The results presented thus far are only particular cases 
of the comprehensive analysis carried out
in this effort, involving several electronic densities, couplings, and temperatures. As
examples of other results obtained in the context of the one-orbital model, in
Fig.~\ref{Figure7}(d) results at $n$=0.7 and in the clean limit are shown. They
also have the peak in the resistivity at the FM transition temperature, to be expected
from the particle-hole symmetry of the model and the $n$=0.3 results. A resistivity peak is also
observed
at $n$=0.2, as shown in Fig.~\ref{Figure7}(f), showing the robustness of the feature. 
However, the particular case $n$=0.5 is special
since in this regime a staggered charge-ordered state is formed and, as a consequence, 
the system is strongly insulating at low temperatures in a wide range 
of electron-phonon couplings $\lambda$ (see Fig.~\ref{Figure7}(e)). This fact was also noticed in Ref.~\onlinecite{re:kumar06}.
Since at this density there is no peak in the resistivity, $n$=0.5 will not be further analyzed here.

\begin{figure*}
\centerline{
\includegraphics[clip,width=13cm]{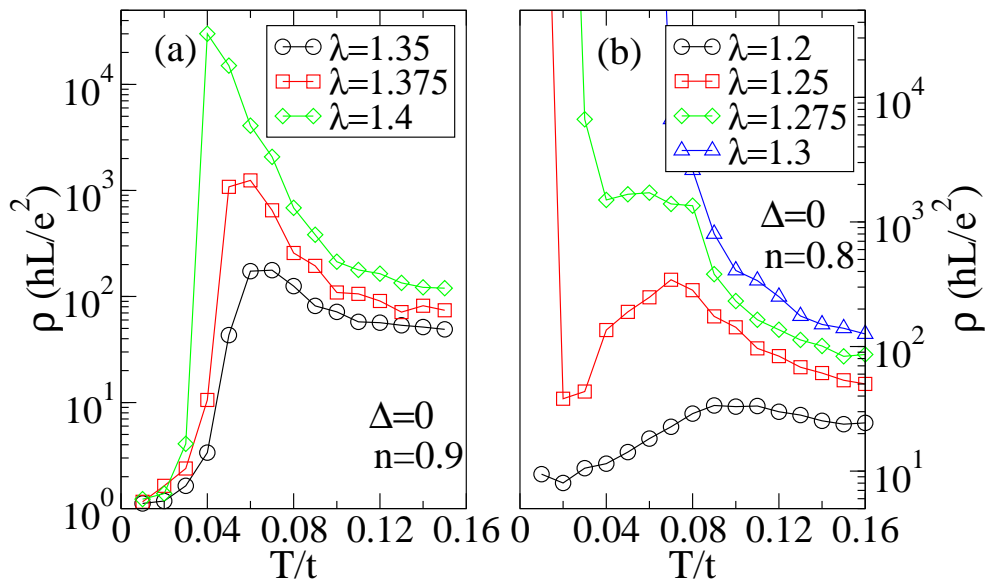}}
\caption{(Color online) Figure to
illustrate the similarity of results obtained using a $6\times6\times6$ lattice (shown)
as compared with the $4\times4\times4$ results discussed before.
(a) Resistivity vs. temperature at $n$=$0.9$ in the clean limit $\Delta$=0,
for the $\lambda$'s indicated.
(b) Resistivity $\rho$ vs. $T$ at $n$=$0.8$ at the $\lambda$'s
indicated and for $\Delta$=0.}
\label{Figure8}
\end{figure*}

To complete the present analysis, size effects have also 
been investigated. It was one of
our purposes to use the standard 
``exact diagonalization'' (ED) method in this study, in order to
avoid considering also issues of
accuracy if an approximate technique would have been employed, 
in addition to the intrinsic subtleties
related with the physics involved in the problem. Moreover, we wanted
to compare our results against those obtained with 
approximate methods carried out
on $8\times8\times8$ lattices \cite{re:kumar06}. The penalization for
using the ED method is that it is 
possible to carry out simulations only on up to 
$6\times6\times6$ clusters and compare
those with the $4\times4\times4$ systems used in the figures discussed so far.
The results are in Fig.~\ref{Figure8}. The existence of the resistivity peak, the overall 
shape of the curves, and the dependence with $\lambda$ are very similar between the two lattices, 
supporting the conclusion that the results are robust and that indeed a
CMR regime has been identified in 
these simulations (and in those reported before in Ref.~\onlinecite{re:kumar06}).

\subsection{The One-electron Problem}

\begin{figure}
\centerline{
\includegraphics[clip,width=8cm,height=7cm]{\mypath{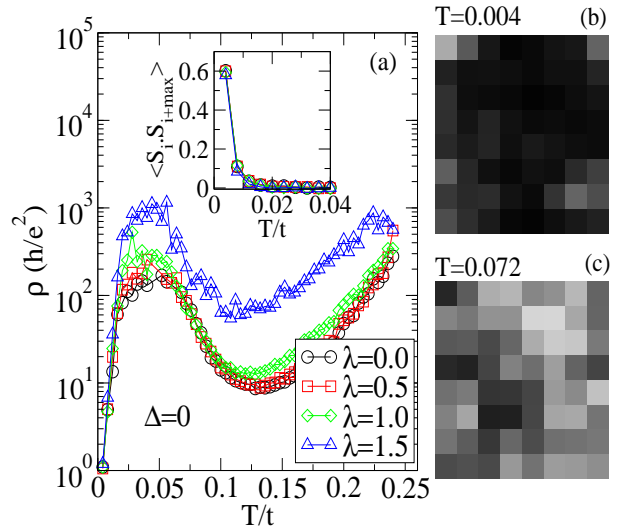}}}
\caption{(Color online) Results obtained in the one electron limit.
(a) Resistivity $\rho$ vs. $T$ at $\Delta=0$ for an $8\times8$ lattice, showing that 
even the $\lambda=0$ results present a peak. The inset contains the spin-spin correlations at the maximum distance;
(b) and (c) are the $\lambda$=0
spatially resolved nearest-neighbor spin-spin correlations $NN(i)$=$\sum_{\langle ij \rangle}
\vec{S}_{i}\cdot\vec{S}_{j}$, where the sum is over the four neighbors $j$ of
site $i$.
The results were obtained 
at $T$=$0.004$ and $T$=$0.072$, 
respectively, namely before and after the resistivity peak. Dark colors denote large values of $NN(i)$,
namely regions where the spins are aligned ferromagnetically.}
\label{Figure9}
\end{figure}

The results reported in the previous sections indicate that the magnitude of
the resistivity peak, namely the ratio between the maximum and minimum resistivities,
increases when reducing the density $n$. In fact, the CMR effect is much
larger at $n$=0.1 than at $n$=0.3. As a consequence, it is natural to wonder if for
the case of just $one$ electron  a peak in the resistivity will also appear.
Our effort is carried out in the grand canonical ensemble, but it is possible
to tune the chemical potential with sufficient accuracy so that 
just one mobile electron is MC simulated.
The results are shown in Fig.~\ref{Figure9}, obtained
on an $8\times8$ cluster. It is remarkable to find that indeed the one-electron problem
has a resistivity vs. temperature curve clearly resembling those of the other
electronic densities. The inset of Fig.~\ref{Figure9}(a) 
shows that ferromagnetism in the classical spins is
obtained in this case as well. In the bulk, likely only a finite-size FM region can be associated
with a single electron, but on a finite small cluster this region can be as large as the
entire system, as it occurs in our case.

An interesting detail of the one-electron study is that the insulating
regime is observed even at $\lambda$=0. This occurs only at this very small electronic
density; at $n$=0.1 or 0.3, a robust value of $\lambda$ is needed to see a similar behavior.
This can be understood as follows. The cluster spin-spin correlations
are sketched in Fig.~\ref{Figure9}(b) at low temperature: here the entire $8\times8$ cluster is 
ferromagnetic in agreement with expectations. However, at higher temperatures, in the ``insulating''
portion of the $\lambda$=0 resistivity curve, there are patches that are FM as well, as shown in
Fig.~\ref{Figure9}(c). This is correlated with charge localized
in the darker regions (not shown). 
\footnote{Actually, for systems without quenched disorder the
average local charge density is always constant. We later
formalize the discussion presented here by introducing the
quantity $\sigma_n$.}
Namely, in the insulating regime there is a ``self-trapping'' of
the electrons that takes place, in the form of a small FM polaron. The lattice does not need to
be distorted to see this curious effect. At temperatures higher
than $T$$\sim$0.125, the resistivity now changes to a metallic state with a more uniform
distribution of charge. The FM polarons at intermediate temperatures have sizes involving several lattice spacings
and, thus,  as $n$ grows it is not surprising that their overlap rapidly renders the system fully metallic.
At densities $n$=0.1 or larger, only with increasing $\lambda$ is that a charge localized regime (with small polarons) 
can be achieved. This issue will be discussed in more detail in the next subsection.

\subsection{Intuitive Understanding of the Results}

To intuitively understand the results, the picture emerging from the one electron
problem is important. It seems that upon cooling a paramagnetic metal first turns
into an insulator via localization of charge (this is the insulating regime of the
resistivity curve) and then, fairly abruptly, a transition to a metallic FM state
occurs. Although we have not calculated the entropy explicitly (this is typically
complicated to do in numerical simulations), we believe that in the range of
$\lambda$'s where this phenomenon occurs, there is a competition between a FM metallic state
and a charge localized (CL) state. The former has lower energy, but the latter has higher
entropy due to the fact that the charge can be localized in a variety of arrangements. For
this reason at high temperature the CL state dominates, but then a crossing to the FM metal
occurs at low temperatures. 

\begin{figure}
\centerline{
\includegraphics[clip,width=9cm]{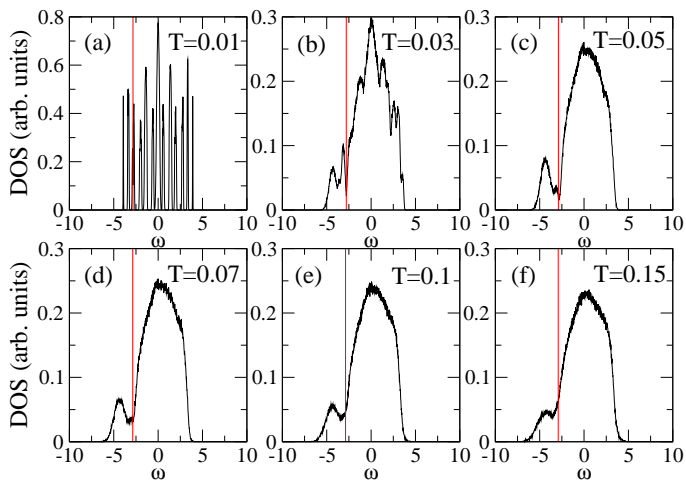}}
\caption{(Color online) (a)-(f) Density-of-states at $\lambda$=$1.5$, $\Delta$=0, using
an $8\times8$ lattice, at the  various temperatures indicated. The red lines (vertical) indicate the location of the
chemical potential such that $n$=0.1 in all the panels.}
\label{Figure10}
\end{figure}

This intuitive picture is compatible with a visual investigation carried out
in this effort.
In the interesting coupling and density regimes, Monte Carlo snapshots 
of the classical spin configurations and electronic density of charge
systematically reveal charge localization in the insulating region. 
This is correlated with the appearance of new structure
in the density-of-states (DOS), as shown
in Fig.~\ref{Figure10}. In this figure, the DOS is shown for the case $n$=0.1 varying
the temperature in the interesting regime identified in Figs.~\ref{Figure5},\ref{Figure6}.
The DOS starts developing a pseudogap (PG) feature at the chemical potential at $T$=0.15.
This PG grows upon cooling and it reaches its maximum depth at the temperature where
the resistivity is maximized. Upon further cooling, the DOS turns into a typical FM 
density-of-states of a finite system, showing multiple spikes \cite{re:dagotto02}.
The presence of a PG in the DOS of a model for manganites
was first observed in Ref.~\onlinecite{re:moreo99}, and our present
results are compatible with those early observations. Clearly, 
the  resistivity peak is unrelated
with Anderson localization that produces a mobility edge in the DOS, 
but not a PG.

\begin{figure}
\centerline{
\includegraphics[clip,width=6.5cm]{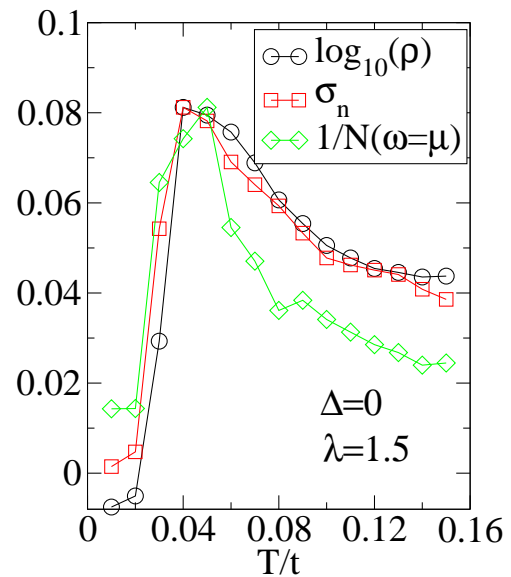}}
\caption{(Color online) Natural logarithm of the resistivity $\rho$ vs. temperature 
in the clean limit, plotted on the same scale with 
$\sigma_n$ (defined in text) and the inverse of the density of states
at the chemical potential, $1/N(\omega=\mu)$, 
working  at $\lambda=1.5$, $n$=0.1, 
and using an $8\times8$ lattice. Both, $\log_{10}(\rho)$ and $1/N(\omega=\mu)$ are normalized to coincide with the maximum of $\sigma_n$. 
Results shown correspond to averages over several independent Monte Carlo runs.
}
\label{Figure11}
\end{figure}

The ideas discussed here related with DOS pseudogaps and charge localization 
can be made more quantitative as follows. In Fig.~\ref{Figure11}, the inverse
of the DOS at the chemical potential is shown, together with the logarithm
of the resistivity. Both quantities show a similar trend with temperature,
and the PG indeed appears
correlated with the behavior of the resistivity. However, note
that the use of the logarithmic scale for $\rho$ indicates that the effect leading
to the PG formation affects much more strongly the transport properties of the
system than others. 
This is typical of a percolative system where small changes in the
electronic distribution can lead to dramatic changes in the transport 
characteristics.


For systems without quenched disorder the average local density
is always constant due
to translational invariance. Therefore, we measure the
localization of the charge by calculating the
error in $n_i$ given by

\begin{equation}
\sigma_{n}^2=\frac{1}{N}\sum_{i}|n_{i}- n  |^{2}.
\end{equation}
This quantity indicates the difference between the actual charge 
$n_i$ at each site and the nominal average density in the full cluster, i.e. $n$. For a 
system with a uniform distribution of charge $\sigma_n$ vanishes. This
indeed occurs at very low temperatures. But for a system with charge localization
then $\sigma_n$ is different from zero, as it occurs at the
resistivity peak. In Fig.~\ref{Figure11}, $\sigma_n$ is plotted 
versus temperature, showing that it follows the behavior of the
resistivity indicating that localization of the
charge is the main reason for the
insulating regime observed above the Curie
temperature. 

\begin{figure*}
\centerline{
\includegraphics[clip,width=15.0cm]{\mypath{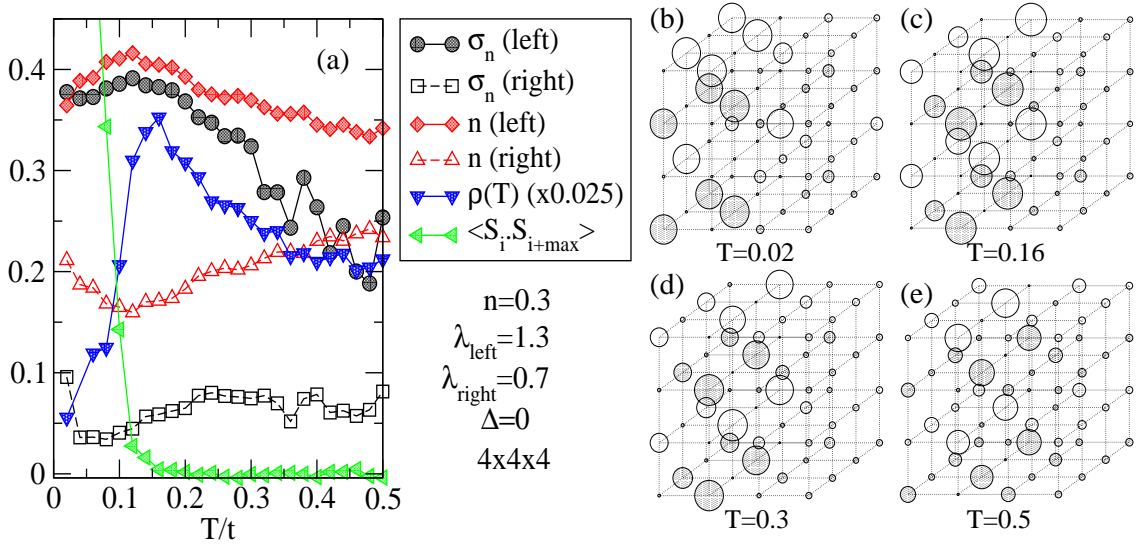}}}
\caption{(Color online) Possible explanation for the
existence of the resistivity peak, using a $4\times4\times4$ and the
couplings and densities indicated. Results shown are for the
artificial  ``left-right'' system described in the text, where
the left (right) of the lattice has a relatively large (small) $\lambda$.
(a) Various quantities (see middle inset) vs. temperature.
The growth of $\sigma_n$ (left) and $n$ (left) with decreasing
temperature is indicative of localization of charge in the
large $\lambda$ region (left).  The trends
in these quantities are very similar to the resistivity in its
insulating range. At the Curie temperature (see spin-spin
correlations), the localization features remains the same but now  the
mobile carriers
can conduct much better than in a paramagnetic spin background.
(b)-(e) Electronic density (proportional to the diameter of the spheres)
at the temperatures indicated.
}
\label{Figurelast}
\end{figure*}

To gain further qualitative understanding for the existence of the peak
in the resistivity, we have studied a special cluster that seems to have common
features with those analyzed thus far. The cluster is a $4\times4\times4$ lattice,
where the 32 sites on the right have a relatively small $\lambda$, i.e. not
strong enough to lead to localization of charge, while the 32 sites on the left
have a large $\lambda$. This allow us to clearly separate in space regions
with and without charge localization. Monte Carlo simulating this system lead
us to the resistivity and spin-spin correlations shown in Fig.~\ref{Figurelast}(a).
The shape is very similar to that of other simulations previously described.
An important point to notice is that the total amount of charge on the left
side of the cluster grows with decreasing temperature and the same does 
$\sigma_n$, the quantity that measures the degree of localization. The increase
of these two quantities with cooling (shown in the figure as well) 
is correlated with the increase of
resistivity, namely with the insulating regime in the resistivity plot. Thus,
it is clear that the insulator portion of the resistance is caused by charge localization. 
Examples of the local densities are in Fig.~\ref{Figurelast}(b-e).

At the temperature where the resistivity turns metallic upon cooling, namely
at the Curie temperature, note that the localization parameters remain approximately the
same as at higher temperatures. Thus, the amount of charge that is localized
does $not$ change dramatically at the metal-insulator transition. What does change is
the spin background and since the conducting properties 
of a ferromagnet and a paramagnet
are very different, then there is a notorious reduction of the resistance below
the Curie temperature. The combination of these two effects leads to the notorious
resistivity peaks found in the Monte Carlo simulations.

\section{TWO-ORBITAL MODEL}

\subsection{Definition}
The two orbitals used in this model arise from the two $e_{\rm g}$ bands that are active at the Mn ions in Mn-oxides,
as extensively discussed before.\cite{re:dagotto01,re:dagotto02,re:dagotto05b}
The Hamiltonian for this model is\cite{re:dagotto01,re:dagotto02,re:dagotto05b}

\begin{eqnarray}
H_{2b}&=&\sum_{\gamma,\gamma',i,\alpha}t^\alpha_{\gamma\gamma'}
{\mathcal S}(\theta_i,\phi_i,\theta_{i+\alpha},\phi_{i+\alpha})
c^\dagger_{i,\gamma}c_{i+\alpha,\gamma'} \nonumber \\ &+&
\lambda\sum_{i}(Q_{1i} \rho_i +  Q_{2i} \tau_{xi} + Q_{3i} \tau_{zi}) \nonumber \\&+&
\sum_{i}\sum_{\alpha=1}^{\alpha=3} D_\alpha Q_{\alpha i}^2,
\label{eq:hamtwobands}
\end{eqnarray}
where the factor that renormalizes the hopping in the $J_{\rm H}$=$\infty$ limit is
\begin{eqnarray}
{\mathcal S}(\theta_i,\phi_i,\theta_{j},\phi_{j})&=&\cos(\frac{\theta_{i}}{2})\cos(\frac{\theta_{j}}{2})
\nonumber \\&+&\sin(\frac{\theta_{i}}{2})\sin(\frac{\theta_{j}}{2})e^{-i (\phi_{i}-\phi_{j})}.
\end{eqnarray}
The parameters $t^\alpha_{\gamma\gamma'}$ are the
hopping amplitudes
between the orbitals $\gamma$ and $\gamma'$ in the direction $\alpha$. In this section, we restrict ourselves
to two dimensions, such that $t^{x}_{aa}=-\sqrt{3}t^{x}_{ab}=
-\sqrt{3}t^{x}_{ba}=3t^{x}_{bb}=1$,
and $t^{y}_{aa}=\sqrt{3}t^{y}_{ab}=\sqrt{3}t^{y}_{ba}
=3t^{y}_{bb}=1$. $Q_{1i}$, $Q_{2i}$ and $Q_{3i}$ are normal
modes of vibration that can be expressed in terms of the oxygen coordinate $u_{i,\alpha}$ as:
\begin{eqnarray}
Q_{1i}&=&\frac {1}{\sqrt{3}} [(u_{i,z}-u_{i-z,z}) + (u_{i,x}-u_{i-x,x}) \nonumber \\ &+&
(u_{i,y}- u_{i-y,y})], \nonumber \\
Q_{2i}&=&\frac {1}{\sqrt{2}} (u_{i,x}-u_{i-x,x}), \nonumber \\
Q_{3i}&=&\frac {2}{\sqrt{6}} (u_{i,z}-u_{i-z,z})
 -\frac{1}{\sqrt{6}} (u_{i,x}-u_{i-x,x}) \nonumber \\&-&\frac{1}{\sqrt{6}} (u_{i,y}- u_{i-y,y}). \nonumber
\end{eqnarray}
Also, $\tau_{xi}=c^{\dagger}_{ia}c_{ib}+c^{\dagger}_{ib}c_{ia}$,
$\tau_{zi}=c^{\dagger}_{ia}c_{ia}-c^{\dagger}_{ib}c_{ib}$, and
$\rho_{i}=c^{\dagger}_{ia}c_{ia}+c^{\dagger}_{ib}c_{ib}$.
The constant $\lambda$ is the electron-phonon coupling
related to the Jahn-Teller distortion of the MnO$_6$
octahedron.\cite{re:rao98,re:moreo99,re:salamon01,re:mathur03,re:ahn04,re:dagotto01,re:dagotto02,re:dagotto05b} Regarding the phononic stiffness, and in
units of $t^{x}_{aa}=1$, the $D_{\alpha}$ parameters are $D_1=1$ and $D_2=D_3=0.5$,
as discussed in previous literature.\cite{re:aliaga03}
The rest of the notation is standard. Note that in the large Hund coupling limit there
is no spin index. The $J_{\rm AF}$ coupling between the localized spins is
neglected, as for the one-orbital model. In some of the results below, a Zeeman term with field
strength $H$ was added.

The main purpose of the numerical study discussed in this section is to
investigate if the two-orbital model for manganites can also produce a
resistivity peak, as observed in the one-orbital case. The study
in this section is presented
with the same caveats as the one-orbital investigation: (i) it is an
important step toward a realistic theoretical description of manganites
since Mn-oxides have two active $e_{\rm g}$ orbitals, but (ii) the model
does not include the coupling $J_{\rm AF}$ which is crucial to generate
the realistic insulating state, with antiferromagnetic and orbital order.
Nevertheless, the observation of features that in several ways resemble experiments
is exciting and at least part of the essence of real materials appears to have
been captured by the models discussed here, even in purely FM regimes.
We are aware that conclusions similar to ours have 
also been reached recently independently
by Kampf and Kumar \cite{re:kampf05}.

\begin{figure}
\centerline{
\includegraphics[clip,width=7cm]{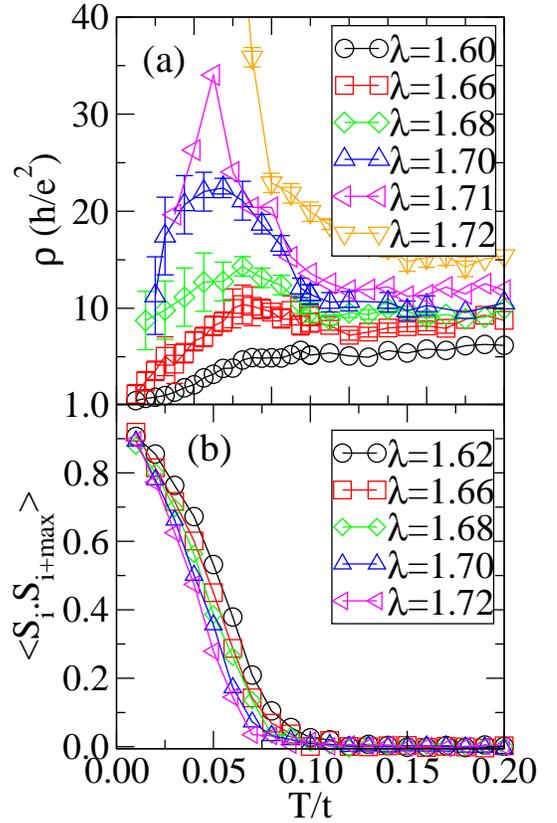}}
\caption{(Color online)
(a) Resistivity vs.~$T$ for various values of $\lambda$, using the
two-orbital model for manganites. The simulation was
performed on an 8$\times$8 lattice with 20 electrons ($n\approx0.3$), using 5,000 MC steps for
thermalization and 5,000 for measurements. These results are in the clean limit $\Delta$=0.
A ferromagnetic arrangement was used as the starting configuration, although tests using paramagnetic
states reveal very similar results. (b)
Spin-spin correlation (at the maximum distance $4\sqrt{2}$ allowed in the studied lattice) vs.~$T$ for various values
 of $\lambda$. Lattice, density, and MC steps are as in (a).
}
\label{Figure12}
\end{figure}

\subsection{Density $n$=0.3}

Typical computational results for the two-orbital model at $n$$\sim$0.3
are shown in Fig.~\ref{Figure12}. 
In the clean limit $\Delta$=0, there is a narrow region of $\lambda$
where a well-defined peak is found in the resistivity. The location
of the peak is correlated with the appearance of ferromagnetic
order, as also shown in the same figure. The systematic tendencies and
behaviors observed in the two-orbital model  simulations are very similar to
those found for the one-orbital model.

\begin{figure}
\centerline{
\includegraphics[clip,width=7cm]{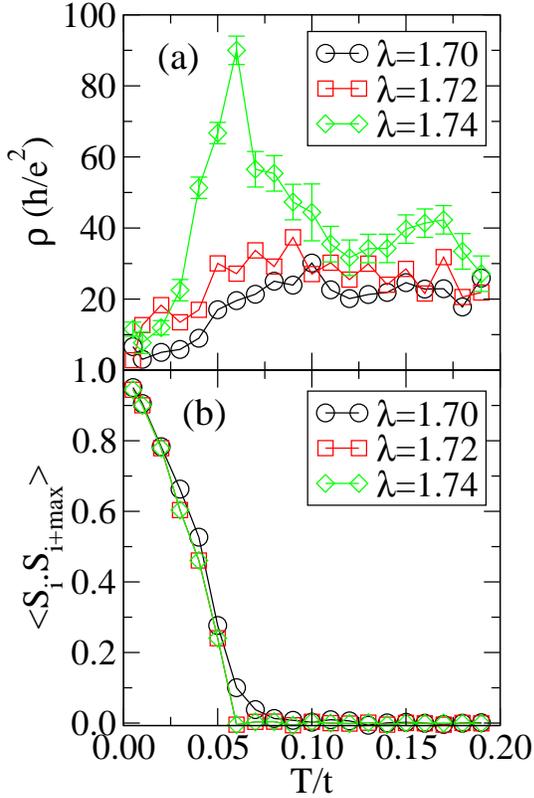}}
\caption{(Color online)  
(a) Resistivity vs. $T$ for a 12$\times$12 lattice in the clean limit $\Delta$=0, 
considering 2 orbitals per site, 44 electrons ($n$$\sim$1/3), $J_{\rm AF}$=0.0, 
and the values of $\lambda$ indicated.
The simulation was carried out starting with a random state at each temperature, 
warming up for 7,500 MC steps per site, followed by 7,500 MC steps for measurements. 
(b) Spin-spin correlation at the largest possible distance ($6\sqrt{2}$)
on a 12$\times$12 lattice, in the clean limit, with the same convention
and parameters as in (a).}
\label{Figure15}
\end{figure}

Finite-size effects do not seem to modify strongly our conclusions, as also found
for the one-orbital case. In Fig.~\ref{Figure15}, results obtained using a $12\times12$
cluster are reported. A resistivity peak is observed in a very similar
range of $\lambda$ as found using the $8\times8$ cluster. Although these results are not
sufficient to fully prove that the behavior found on finite lattice survives the bulk
limit, they are very suggestive: in two and three dimensional lattices, for a wide
range of electronic densities, for a variety of lattice sizes, with and
without quenched disorder, and both for the one- and
two-orbital models 
the resistivity peak is present in the study of charge
transport.

\begin{figure}
\centerline{
\includegraphics[clip,width=7cm]{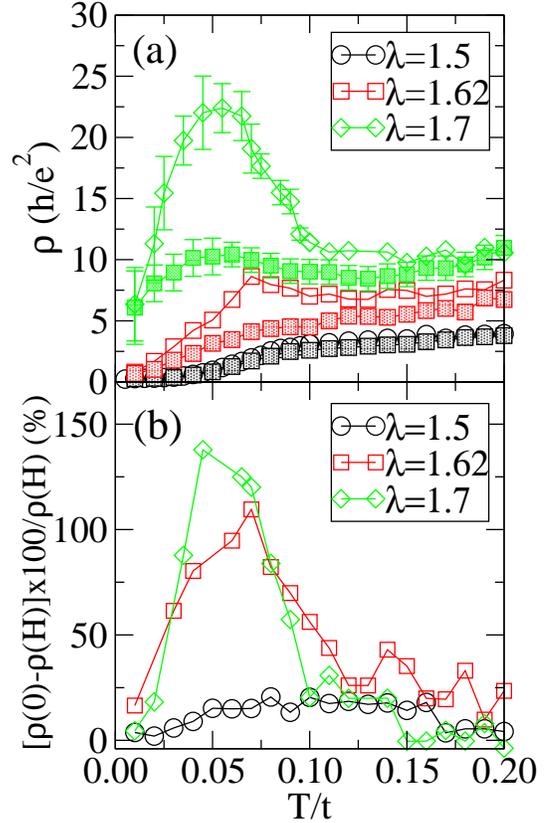}}
\caption{(Color online)
(a) Resistivity vs.~$T$ for several $\lambda$'s (indicated). Results were obtained
 with (filled symbols) and without (open symbols) a  magnetic field $H$=0.1. The simulation was
performed on an 8$\times$8 lattice, with 20 electrons ($n\approx0.3$), 5,000 thermalization MC
steps followed by  5,000 MC steps for measurements.
(b) Magneto-resistance 
(defined as (($\rho(0)$-$\rho(H)$)/$\rho(H)$)$\times$100 vs.~$T$ 
for the same parameters as in (a).}
\label{Figure13}
\end{figure}

This resistivity peak in the two-orbital 
model is also drastically affected by
relatively small magnetic fields, as observed for the one-orbital case. Typical
results are in Fig.~\ref{Figure13}. The region the most affected by the
magnetic fields is where the maximum in the resistivity is observed, 
as expected. The magnitude of the magnetoresistance effect shown in 
the figure is comparable to the numbers found for the one-orbital case 
at similar electronic densities (see Fig.~\ref{Figure3}).

\begin{figure}
\centerline{
\includegraphics[clip,width=7cm]{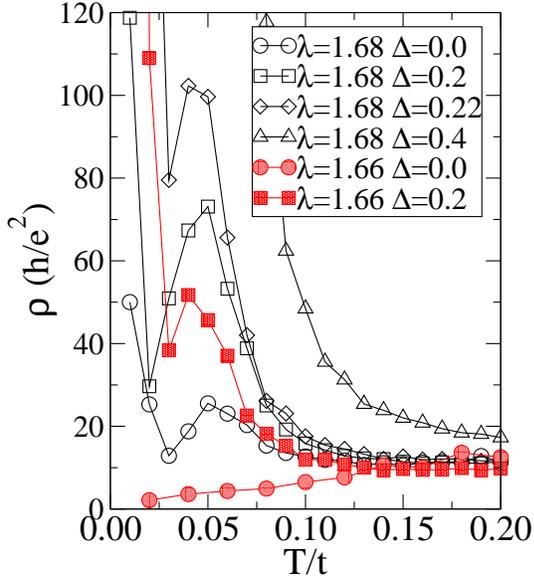}}
\caption{(Color online)
Resistivity vs.~$T$ for various values of $\lambda$ and 
with and without quenched disorder, as indicated. The figure shows the enhancement of
the resistivity peak with increasing $\Delta$.
The lattice size, MC steps, and various
parameters are as in Fig.~\ref{Figure13}.}
\label{Figure14}
\end{figure}

It is also important to discuss the influence of
quenched disorder. Typical results are in Fig.~\ref{Figure14}, where the resistivity
vs. $T$ is shown both with and without disorder. As anticipated from recent
previous investigations \cite{re:kumar06},
and from the one-orbital study in this manuscript, it was observed that 
introducing quenched disorder enhances substantially
the features found in the clean limit. For instance, at $\lambda$=1.66, the $\Delta$=0 curve does not
show a resistivity peak, but this feature is generated at $\Delta$=0.2 and the same $\lambda$. In cases
where the resistivity
already has a peak in the clean limit, this structure is enlarged with increasing $\Delta$.

\subsection{Density $n$=0.1}

\begin{figure}
\centerline{
\includegraphics[clip,width=7cm]{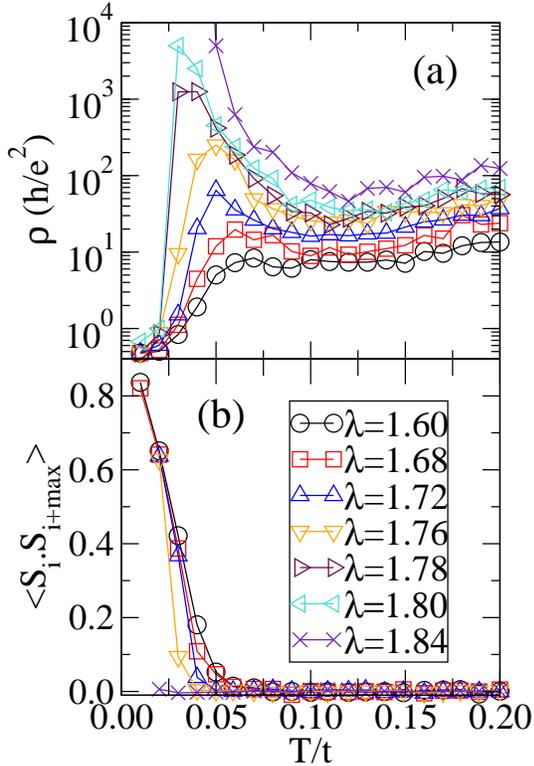}}
\caption{(Color online)
Results for the two-orbital model at $n$=0.1 and the $\lambda$'s
indicated. (a)
Resistivity vs. temperature using an $8\times8$ lattice, 5,000 MC
steps for thermalization and a similar number of measurements.
Note how sharp is the low temperature transition from low to high
resistance.
(b) Spin-spin correlation at the maximum distance. 
}
\label{Figure16}
\end{figure}

In Fig.~\ref{Figure16}, Monte Carlo results at $n$=0.1 are
presented for the two-orbital model. 
The behavior of the resistivity is very similar
to what was observed for the one-orbital case
(see Figs.~\ref{Figure8}(a) and \ref{Figure6}(a)), namely a clearly
defined peak is observed, and a sharp (likely first order) 
transition in the resistance
occurs upon cooling through the Curie temperature.
At this electronic density, the changes in resistance upon heating
or cooling are much larger than at other densities such as $n$=0.3.

Overall, it is clear that the models with one and two orbitals
behave fairly similarly, and
the existence of a peak in the resistivity is a robust result of this effort
and previous Monte Carlo simulations \cite{re:kumar06}.

\section{Conclusions}

The research effort discussed in this paper reached several goals.
First, it confirmed recent reports by other groups \cite{re:verges02,re:kumar06} regarding
the existence of a large peak in the resistivity vs. temperature 
for the one-orbital model for manganites, including a robust electron-phonon coupling.
This confirmation is interesting since the results of the 
previous\cite{re:kumar06} and current efforts
were obtained using different techniques to
estimate transport properties, and also with different methods to simulate
the one-orbital model. Second, a comprehensive 
analysis of the influence of couplings, quenched disorder strength, and 
electronic density
was here described.  This includes the case of just one electron 
on an otherwise carrier empty lattice,
in the presence of classical $t_{\rm 2g}$ spins. 
This one-electron problem also presents a large resistance peak
when varying the temperature. A very simple explanation for the behavior of these
systems was discussed, based on a competition between tendencies to charge
localization and ferromagnetism. 

Finally, the present effort also includes a study of the two-orbital 
model for manganites. The overall conclusion
is that its behavior is similar to that of the one-orbital model. 
Since these results are themselves also 
similar to experiments, our effort and those of other groups provide 
evidence that the theoretical studies that focus
on the regime of competition between a metal and an insulator are on the 
right track toward a full explanation of the CMR phenomenon. 
Both with one and two orbitals, quenched disorder is important to enlarge the
magnitude of the effects and broaden its range in parameter space, 
thus avoiding the fine tuning of couplings needed in the clean limit. 

In the present and related investigations \cite{re:verges02,re:kumar06} 
both the metal and the insulator at low temperatures
are ferromagnetic, and they differ only in the arrangement of charge (extended vs. localized 
character). The next level of sophistication of the simulations of manganite models must address 
the competition between different magnetic orders. Although the current results 
have a remarkable resemblance with several experiments, it is known that the largest effects 
in real manganites
occur when a antiferromagnetic/charge/orbital ordered state competes with
the ferromagnetic metal. To achieve this final goal, 
the coupling $J_{\rm AF}$ must be incorporated in the
investigations. This will require levels of numerical
accuracy higher than in the present effort, due to the competition of very different
states that typically lead to metastabilities and long thermalization times. Results will hopefully
be presented in the near future.

\section{ACKNOWLEDGMENTS}
We thank S. Kumar and A. Kampf for useful conversations. 
This work is supported in part by the LDRD
program at ORNL and by the NSF grant DMR-0443144.
Most of the computational work in this effort was performed 
at the supercomputing facilities of the Center for Computational Science at 
the Oak Ridge National Laboratory (ORNL), managed by UT-Battelle, LLC, for the 
U.S. Department of Energy under Contract DE-AC05-000R22725.
We also acknowledge the help of J. A. Verg\'{e}s in the study of conductances. 
This research used the SPF computer program and software toolkit developed at ORNL (http://mri-fre.ornl.gov/spf).

\bibliography{paper5}

\end{document}